\documentclass[useAMS,usenatbib]{mn2e}
\usepackage[letterpaper]{geometry}
\usepackage{times}
\usepackage{amsfonts,amsmath,amssymb}
\usepackage{graphicx}
\usepackage{epstopdf}

\DeclareMathOperator{\erf}{erf}
\DeclareMathOperator{\sinc}{sinc}

\title[Gravitational wave source timing parallax]{Pulsar timing array observations of gravitational wave source timing parallax}
\author[X. Deng and L. S. Finn] {Xihao Deng${}^1$
and 
Lee Samuel Finn${}^{1,2}$\\
${}^{1}$Department of Physics, The Pennsylvania State University, University Park PA 16802\\
${}^{2}$Department of Astronomy and Astrophysics, The Pennsylvania State University, University Park PA 16802}
\begin{document}
\date{}
\pagerange{\pageref{firstpage}--\pageref{lastpage} \pubyear{2010}}
\pubyear{}

\maketitle
\label{firstpage}

\begin{abstract}
Pulsar timing arrays act to detect gravitational waves by observing the small, correlated effect the waves have on pulse arrival times at Earth. This effect has conventionally been evaluated assuming the gravitational wave phasefronts are planar across the array, an assumption that is valid only for sources at distances $R\gg2\pi{}L^2/\lambda$, where $L$ is physical extent of the array and $\lambda$ the radiation wavelength.  In the case of pulsar timing arrays (PTAs) the array size is of order the pulsar-Earth distance (kpc) and $\lambda$ is of order pc. Correspondingly, for point gravitational wave sources closer than $\sim100$~Mpc the PTA response is sensitive to the source parallax across the pulsar-Earth baseline. Here we evaluate the PTA response to gravitational wave point sources including the important wavefront curvature effects. Taking the wavefront curvature into account the relative amplitude and phase of the timing residuals associated with a collection of pulsars allows us to measure the distance to, and sky position of, the source. 


\end{abstract}


\section{Introduction}

It has been just over thirty years since \citet{sazhin:1978:ofd} and \citet{detweiler:1979:ptm} showed how passing gravitational waves cause disturbances in pulsar pulse arrival times that could be used to detect or limit gravitational wave signal strength, and  twenty years since \citet{foster:1990:cpt} proposed using correlated timing residuals of a collection of pulsars --- i.e., a \emph{pulsar timing array} (PTA) --- to achieve greater sensitivity. Thought of as a gravitational wave detector a PTA is large: its size $L$ (on order kpc) is much greater than the gravitational radiation wavelength scale $\lambda$ (on order pc) that we use it to probe. The conventional approximation of gravitational waves as plane-fronted applies only for sources at distances $R\gg{}L^2/\lambda$: for less distant sources the curvature of the gravitational radiation phasefronts contribute significantly to the detector response.  Here we evaluate the response of a PTA to discrete gravitational wave sources at distances close enough that gravitational wave phasefront curvature is important and describe how PTA observations of such sources may be used to measure the source distance and location on the sky. 


Pulsar timing currently provides the best observational evidence for gravitational waves: as of this writing the orbital decay of the binary pulsar system PSR~B$1913+16$, induced by gravitational wave emission and measured by timing measurements, is the most conclusive observational evidence for gravitational waves \citep{taylor:1979:mog,weisberg:2005:rbp}. This evidence is often described, without any intent of disparagement, as \emph{indirect} since it is the effect of gravitational wave \emph{emission} --- i.e., the effect of the emission on the emitter --- that is observed.\footnote{Indirect observation is enough to inaugurate the field of gravitational wave astronomy: i.e., the study of the physical universe through the application of the laws and theories to astronomical observations involving gravitational waves. See, for example, \citet{finn:2002:bmo,sutton:2002:bgm}, which use the effect of gravitational wave emission on several binary pulsar systems to bound the mass of the graviton.} The effect of the gravitational radiation, independent of its source, on another system --- a \emph{detector} --- itself remains to be observed. 

As proposed by \citet{sazhin:1978:ofd} and \citet{detweiler:1979:ptm} precision pulsar timing can also be used to detect gravitational waves \emph{directly.} Gravitational waves passing between the pulsar and Earth--- i.e., temporal variations in the space-time curvature along the pulsar-Earth null geodesics --- change the time required by successive pulses to travel the path from pulsar to Earth.  Since successive pulses are emitted from the pulsar at regular intervals the effect of passing gravitational waves are apparent as irregularities in the observed pulse arrival times. Through the use of precision timing a collection of especially stable pulsars may thus be turned into a galactic scale gravitational wave detector. 

Timing of individual pulsars has been used to bound the gravitational wave power associated with stochastic gravitational waves of cosmological origin \citep{romani:1983:ulo,kaspi:1994:hto,lommen:2002:nlo}; timing residual correlations among pulsar pairs have been used to bound the power in an isotropic stochastic gravitational wave background \citep{jenet:2006:ubo}; the absence of evidence for gravitational waves in the timing of a single pulsar has been used to rule-out the presence of a proposed supermassive black hole binary in 3C~66B \citep{jenet:2004:cpo}; observations from multiple pulsars have been used to place limits on periodic gravitational waves from Sag~$\mathrm{A}^*$ \citep{lommen:2001:upt}; and timing analyses of individual pulsars have been combined to place a sky-averaged constraint on the merger rate of nearby black-hole binaries in the early phases of coalescence \citep{yardley:2010:sop}. Additionally, analyses of pulsar timing data have been proposed to search for the gravitational wave ``memory'' effect \citep{van-haasteren:2010:gma,seto:2009:sfm,pshirkov:2010:ogw} and more general gravitational wave bursts \citep{finn:2010:dla}, which might be generated during, e.g., the periapsis passage of a highly eccentric binary system \citep{sesana:2010:scm} or by cosmic string cusps or kinks \citep{damour:2001:gwb, siemens:2007:gsb,leblond:2009:gwf}.

Analyses focused on discovering or bounding the intensity of a stochastic gravitational wave ``background'' may rightly resolve the radiation into a superposition of plane waves with density $\partial^3\widetilde{\mathbf{h}}/\partial f\partial^2\Omega$, where $\mathbf{h}(t,\vec{x})$ is the gravitational wave strain and $(f,\Omega)$ denotes a plane wave component at frequency $f$ and propagating in direction $\Omega$. Analyses focused on the discovery of gravitational wave point sources --- e.g., supermassive black hole binaries --- must take proper account of the curvature of the radiation phase-front, owing to the finite distance between the source and the detector, over the detector's extent. When detector extent, measured in units of the radiation wavelength, is greater than the distance to the source, measured in units of detector extent, then curvature of the radiation wavefront over the detector contributes significantly to the detector response to the incident radiation. 

In \S\ref{sec:resp} we evaluate the pulse arrive-time disturbance associated with a spherically-fronted gravitational wave traversing a pulsar-Earth line-of-sight and compare it with the response to the same wave in the plane-wave approximation. In \S\ref{sec:disc} we describe how, owing to their sensitivity to gravitational radiation phasefront curvature, PTA observations of point sources can be used to measure or place lower bounds on the distance to the source distance and, more surprisingly, the distance to the array pulsars. Finally we summarize our conclusions in \S\ref{sec:concl}. 

\onecolumn
\section{Pulsar timing residuals from spherically-fronted gravitational waves}\label{sec:resp}
\subsection{Timing residuals}\label{sec:residuals}
Focus attention on the electromagnetic field associated with the pulsed emission of a pulsar and denote the field phase, at the pulsar, as $\phi_0(t)$. We are interested in the time-dependent phase $\phi_0(t)$ of the electromagnetic field associated with the pulsed emission measured at an Earth-based radio telescope, which we write as
\begin{subequations}
\begin{equation}
\phi(t) = \phi_0[t-L-\tau_0(t)-\tau_{\mathrm{GW}}(t)]
\end{equation}
where
\begin{align}
\tau_0 &= 
\left(\begin{array}{l}
\text{Corrections owing exclusively to the spatial motion of the Earth}\\
\text{within the solar system, the solar system with respect to the pulsar,}\\
\text{and electromagnetic wave propagation the interstellar medium}
\end{array}\right)\\
\tau_{\mathrm{GW}} &= 
\left(\begin{array}{l}
\text{Corrections owing exclusively to $\mathbf{h}(t,\mathcal{X})$}
\end{array}\right)\\
L &= \left(\text{Earth-pulsar distance}\right). 
\end{align}
\end{subequations}
(Note that we work in units where $c=G=1$.) 

In the absence of gravitational waves $\tau_{\mathrm{GW}}$ vanishes and the phase front $\phi_0(t)$ arrives at Earth at time $t_\oplus(t) = t+L+\tau_0(t)$. In the presence of a gravitational wave signal the phase front arrives at time $t_\oplus(t)+\tau_{\mathrm{GW}}(t)$; thus, $\tau_{\mathrm{GW}}$ is the gravitational wave timing residual. Following \citet{finn:2009:roi} Eqs.~(3.26) and (3.12e), the arrival time correction $\tau_{\mathrm{GW}}(t)$ is 
\begin{subequations}\label{eq:tauGW}
\begin{align}
\tau_{\mathrm{GW}}(t)
&= -\frac{1}{2}\hat{n}^l\hat{n}^m\mathcal{H}_{lm}(t)
\end{align}
where $\mathcal{H}_{lm}$ is the integral of the transverse-traceless metric perturbation over the null geodesic ranging from the pulsar to Earth:
\begin{align}
\mathcal{H}_{lm}(t) &= L\int_{-1}^0
h_{lm}\left(t+L\xi,\mathcal{P}-L(1+\xi)\hat{n}\right)\,d\xi
\label{eq:scrH}\\
\mathcal{P} &= \left(\text{Pulsar location}\right)\\
\hat{n} &= \left(\text{Unit vector pointing from Earth to pulsar}\right).
\end{align}
\end{subequations}

\subsection{Gravitational waves from a compact source}\label{sec:gravWaves}
At a point $\mathcal{X}$ in the perturbative regime far from the source the transverse-traceless gauge gravitational wave metric perturbation associated with radiation from a compact source may be written \citep{finn:1985:gwf} 
\begin{subequations}\label{eq:sphH}
\begin{align}
h_{lm}(t,\mathcal{X}) &= 
\int_{-\infty}^{\infty} \widetilde{h}_{lm}(f,\hat{k}) e^{-2\pi ift}df\\
&=\frac{1}{\left|\mathcal{X}-\mathcal{S}\right|}\int_{-\infty}^{\infty} \left[\widetilde{A}_{lm}(f,\hat{k}) e^{2\pi if|\mathcal{X}-\mathcal{S}|}\right]e^{-2\pi ift}df
\end{align}
where 
\begin{align}
\mathcal{S} &= \left(\text{source location}\right)\\
\hat{k} &= \frac{\mathcal{X}-\mathcal{S}}{\left|\mathcal{X}-\mathcal{S}\right|}
= \left(\text{unit vector in direction of wave propagation at $\mathcal{X}$}\right)
\end{align}
\end{subequations}
The Fourier coefficient functions $\widetilde{A}_{lm}(f,\hat{k})$ are everywhere symmetric, traceless and transverse with respect to $\mathcal{X}-\mathcal{S}$: i.e., 
\begin{align}
\widetilde{A}_{lm}(f,\hat{k}) &= \widetilde{A}_{+}(f,\hat{k})\mathbf{e}^{(+)}_{lm} + \widetilde{A}_{\times}(f,\hat{k})\mathbf{e}^{(\times)}_{lm}
\end{align}
where $\mathbf{e}^{(+)}_{lm}(\hat{k})$ and $\mathbf{e}^{(\times)}_{lm}(\hat{k})$ are the usual transverse-traceless gravitational wave polarization tensors for waves traveling in direction $\hat{k}$. 

\subsection{Response function}
With expression \ref{eq:sphH} for $h_{lm}$ we can evaluate $\mathcal{H}_{lm}$ in the 
Fourier domain:
\begin{subequations}
\label{eq:tildeScrH}
\begin{align}
\widetilde{\mathcal{H}}_{lm}(f) &= 
L\int_{-1}^0
\widetilde{h}_{lm}\left(f,\mathcal{P}-L(1+\xi)\hat{n}\right)
e^{-2\pi i fL\xi}
\,d\xi\\
&= 
L\int_{-1}^0 
\frac{\widetilde{A}_{lm}\left(f,\hat{k}(\xi)\right)}{\left|\mathcal{P}-L(1+\xi)\hat{n}-\mathcal{S}\right|}
\exp\left[{2\pi i f\left(-L\xi+|\mathcal{P}-L(1+\xi)\hat{n}-\mathcal{S}|\right)}\right]
\,d\xi
\end{align}
\end{subequations}

\begin{figure}
\begin{center}
\includegraphics[width=4in]{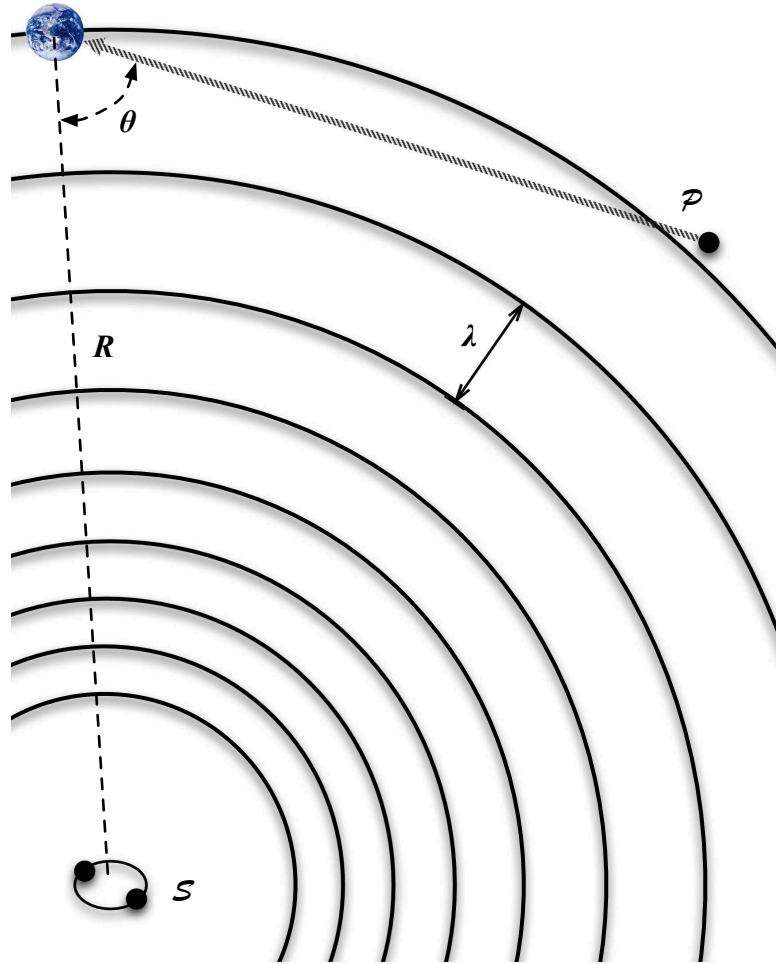}
\end{center}
\caption{Gravitational wave source (depicted as a binary system), Earth and pulsar. The gravitational wave phasefronts intersect the pulsar-Earth path, traveled by the electromagnetic pulses, in a manner that depends on the distance to the source and the source-Earth-pulsar angle $\theta$.}\label{fig:geometry}
\end{figure}

Figure \ref{fig:geometry} describes the Earth-pulsar-source geometry: note we define $R$ to be the Earth-source distance and $\theta$ the angle between the Earth-pulsar and Earth-source lines-of-sight. Along the path traversed by the pulsar's electromagnetic pulse phase fronts ($\mathcal{P}-L\xi\hat{n}$ for $-1\leq\xi\leq0$) we may write
\begin{subequations}
\begin{align}
\hat{k}(\xi) &= \frac{\mathcal{P}-L(1+\xi)\hat{n}-\mathcal{S}}{\left|\mathcal{P}-L(1+\xi)\hat{n}-\mathcal{S}\right|}\\
r(\xi) &= \left(\text{distance between gravitational wave phase front and source}\right)\nonumber\\
&= \left|\mathcal{P}-L(1+\xi)\hat{n}-\mathcal{S}\right|
= R\left[1+2\frac{L\xi}{R}\cos\theta+\left(\frac{L\xi}{R}\right)^2\right]^{1/2}.
\label{eq:r}
\end{align}
\end{subequations}

From Equations \ref{eq:tildeScrH} and \ref{eq:r} we note that the time of arrival disturbance owing to a passing gravitational wave depends on the dimensionless quantities $\pi f L$ and $L/R$. PTA pulsar distances $L$ are all on-order kpc. Gravitational waves detectable via pulsar timing observations have frequencies greater than the inverse duration of the observational data set and less than the typical sampling period: i.e., $10^{-5}~Hz\gtrsim f\gtrsim10^{-9}$~Hz. Strong sources of gravitational waves (e.g., supermassive black hole binaries and triplets) in this band are all expected to be extragalactic, with significant numbers within on order 100~Mpc \citep{sesana:2010:gwa}. The scales of interest are thus either very large or very small: 
\begin{subequations}\label{eq:numerology}
\begin{align}
\epsilon &= 10^{-5}\left(\frac{L}{1\,\text{kpc}}\right)\left(\frac{100\,\text{Mpc}}{R}\right)\\
\pi fL &= 10^4\,\left(\frac{f}{1\,\text{yr}^{-1}}\right)\left(\frac{L}{1\,\text{kpc}}\right)\\
\pi fL\epsilon &= 10^{-1}\,\left(\frac{f}{1\,\text{yr}^{-1}}\right)\left(\frac{L}{1\,\text{kpc}}\right)^2\left(\frac{100\,\text{Mpc}}{R}\right).
\end{align}
\end{subequations}
Taking advantage of these scales the Fourier transform $\widetilde{\mathcal{H}}_{lm}(f)$ may be expressed
\begin{subequations}
\begin{align}
\widetilde{\mathcal{H}}_{lm}(f) &= 
L\int_{-1}^0
\widetilde{h}_{lm}\left(f,\mathcal{P}-L(1+\xi)\hat{n}\right)
e^{-2\pi i fL\xi}
\,d\xi\\
&= 
L\int_{-1}^0 
\frac{\widetilde{A}_{lm}\left(f,\hat{k}(\xi)\right)}{\left|\mathcal{P}-L(1+\xi)\hat{n}-\mathcal{S}\right|}
e^{2\pi i f\left(-L\xi+|\mathcal{P}-L(1+\xi)\hat{n}-\mathcal{S}|\right)}
\,d\xi\\
&= 
\widetilde{A}_{lm}(f,\hat{k}_0) 
\frac{\exp\left[\pi i fR\left(2-\frac{1-\cos\theta}{1+\cos\theta}\right)\right]}{\sqrt{fR}\sin\theta}
\frac{e^{-i\pi/4}}{2}
\left.
\erf\left[e^{i\pi/4}x\right]
\right|^{\sqrt{\pi fR}\frac{\epsilon(1+\cos\theta)+1}{1+\cos\theta}\sin\theta}_{x=\frac{\sqrt{\pi fR}\sin\theta}{1+\cos\theta}} \nonumber\\
&\qquad{}
\times\left[1+
\mathcal{O}(\epsilon)
+\mathcal{O}\left(\epsilon\frac{\partial\log\tilde{A}}{\partial{\hat{k}}}\cdot\hat{n}\right)
\right]
\end{align}
\end{subequations}

Except when $\theta\leq\left(\pi fR\right)^{-1/2}\ll1$ the error function arguments are always large in magnitude. Taking advantage of the asymptotic expansion of $\erf(z)$ about the point at infinity,
\begin{align}
\erf(z) \sim 1 - \frac{\exp\left(-z^2\right)}{z\sqrt{\pi}}
\end{align} 
we may thus write
\begin{subequations}\label{eq:tau}
\begin{align}
\widetilde{\tau}_{\text{gw}} &= -\frac{1}{2}\hat{n}^l\hat{n}^m\widetilde{\mathcal{H}}_{lm}(f) \\
&= \widetilde{\tau}_{\mathrm{pw}} + \widetilde{\tau}_{\mathrm{cr}}
\end{align}
where
\begin{align}
\widetilde{\tau}_{\mathrm{pw}}
&=
\left(
\text{plane wave approximation timing residual}
\right)\nonumber\\
&= 
2\widetilde{\mathcal{A}}(f,\hat{k}_0)
\exp\left[2\pi ifL\sin^2\frac{\theta}{2}\right]
\sinc\left(2\pi fL\sin^2\frac{\theta}{2}\right)
\\
\widetilde{\tau}_{\mathrm{cr}}
&= 
\left(
\text{phasefront curvature correction to plane wave timing residual}
\right)\nonumber\\
&=
\epsilon\left(1+\cos\theta\right)\widetilde{\mathcal{A}}(f,\hat{k}_0)
\exp\left[
\pi ifL\left(4\sin^2\frac{\theta}{2}+\frac{\epsilon}{2}\sin^2\theta\right)
\right]
\sinc\left[\frac{\pi fL\epsilon}{2}\sin^2\theta\right]\\
\widetilde{\mathcal{A}}(f,\hat{k}_0) &= 
\frac{i\epsilon}{4}\left[\hat{n}^l\hat{n}^m\widetilde{A}_{lm}(f,\hat{k}_0)\right]\exp\left[2\pi i fR\right]
\end{align}
and
\begin{align}
\sinc(x) &=\left(\begin{array}{l}\text{Unnormalized}\\\text{sinc function}\end{array}\right) 
= \frac{\sin(x)}{x}.
\end{align}
\end{subequations}
These expressions are valid for all $\pi fL\gg1$. 

\twocolumn
\section{Discussion}\label{sec:disc}
The Fourier amplitude and phase of the gravitational wave timing residual $\widetilde{\tau}_{\text{GW}}$ for any particular pulsar depends on the pulsar's position and distance relative to the position and distance of the gravitational wave source. Correspondingly, observation in three or more pulsars of the gravitational wave timing residuals associated with a single source can, in principle, be used to infer the three-dimensional source position and location on the sky. In this section we discuss the conditions under which sufficiently accurate measurements of the $\widetilde{\tau}_{\text{GW}}$ are possible and provide a proof-of-principle demonstration of how they may be combined to determine the sky location and distance to a gravitational wave source. 

\subsection{Plane wave approximation timing residuals}\label{sec:planeWave}
Focus attention first on $\widetilde{\tau}_{\text{pw}}$: the plane-wave approximation to $\widetilde{\tau}_{\text{gw}}$.  In order to relate the angular location of a gravitational wave source to the relative amplitude and phase of the $\widetilde{\tau}_{\text{pw}}$ for two or more PTA pulsars the uncertainties $\sigma_L$ in the Earth-pulsar distances, and $\sigma_{\theta}$ in pulsar sky locations, must satisfy
\begin{subequations}
\label{eq:sigmaLT}
\begin{align}
\sigma_L&\lesssim 2\,\text{pc}
\left(\frac{0.1\,\text{yr}^{-1}}{f}\right)
\left(\frac{1/2}{\sin^2(\theta/2)}\right)\\
\sigma_\theta&\lesssim
{2.5'}\left(\frac{0.1\,\text{yr}^{-1}}{f}\right)\left(\frac{\text{kpc}}{L}\right)\left(\frac{\pi/4}{|\sin\theta|}\right).
\end{align}
\end{subequations}
The constraint on pulsar angular position accuracy is not a severe one; however, the constraint on pulsar distance accuracy is quite significant. For this reason previous pulsar timing searches for periodic gravitational waves \citep{yardley:2010:sop,jenet:2004:cpo,lommen:2001:upt,lommen:2001:pmt} have sought only to identify an anomalous periodic contribution to pulse arrival times, avoiding explicit use of the amplitude of $\widetilde{\tau}_{\text{gw}}$. 

Improvements in instrumentation and timing techniques are making possible pulsar distance measurements of the accuracy and precision necessary to relate the $\widetilde{\tau}_{\text{pw}}$ to the gravitational wave source angular location. For example, in recent years a succession of VLBI observations have established the distance to J$0437-4715$ with an accuracy of $\pm1.3\,\text{pc}$ \citep{verbiest:2008:pto,deller:2008:ehp,deller:2009:pva}; correspondingly, for J$0437-4715$ predictions of the amplitude and phase of $\widetilde{\tau}_{\text{pw}}$ can be made and take part in the analysis of timing data to search for gravitational waves at frequencies $f\lesssim0.1\,\text{yr}$. Accurate pulsar distance measurements have been identified as a crucial element of a continuing program to use precision pulsar timing and interferometry to test gravity and measure neutron star properties \citep{cordes:2009:tog}; correspondingly, we can reasonably expect that the number of pulsars with high-precision distances will increase rapidly over the next several years. 

Looking toward the future, observations with the SKA \citep{smits:2009:psa} using existing timing techniques are expected to be capable of measuring timing parallax distances to millisecond pulsars at 20~kpc with an accuracy of 20\% \citep{tingray:2010:har}. When measured by VLBI or timing parallax the fractional uncertainty in the pulsar distance is equal to the fractional uncertainty in the semi-annual pulse arrival time variation owing to the curvature of the pulsar's electromagnetic phasefronts: i.e., 
\begin{subequations}
\begin{align}
\frac{\sigma_L}{L} &= \frac{\sigma_{\tau}}{\Delta\tau}
\end{align}
where $\sigma^2_{\tau}$ is the timing noise power in the bandwidth $T^{-1}$ (for $T$ the observation duration) about a frequency of $2/yr$ and
\begin{align}
\Delta\tau &= \left(\frac{\text{au}}{2c}\right)\left(\frac{\text{au}}{L}\right)
= 1.2\,\mu\text{s}\,\left(\frac{\text{kpc}}{L}\right)
\end{align}
\end{subequations}
is the peak-to-peak variation in the pulse arrival time residual owing to the curvature of the pulsar's electromagnetic phasefronts. Correspondingly, a 20\% accuracy measurement of the distance to a pulsar at 20~kpc corresponds to an 0.5\% accuracy measurement of the distance to a pulsar of the $\sigma_{\tau}$ at 500~pc: i.e., 2.5~pc. 

This estimate of $\sigma_L=2.5$~pc for a pulsar at 500~pc assumes constant signal-to-noise and, correspondingly, constant timing precision $\sigma_{\tau}$. Closer pulsars will have greater fluxes and, correspondingly, better timing precision; consequently we may regard this as an upper bound on the distance measurement error $\sigma_L$. If we assume that $\sigma_{\tau}^{-2}$ is proportional to the pulsar energy flux then 
\begin{subequations}\label{eq:sigmaL}
\begin{align}
\sigma_{\tau} &= 0.6\,\text{ns}\,\left(\frac{L}{\text{kpc}}\right)\\
\sigma_L &= 0.5\,\text{pc}\,\left(\frac{L}{\text{kpc}}\right)^3; 
\end{align}
\end{subequations}
i.e., if the only barrier to improved timing precision is signal-to-noise then sub-parsec precision distance measurements should be possible in the SKA era for pulsars within a kpc. 

In practice other contributions to the noise budget will eventually limit the timing precision. Among these the most important are intrinsic pulsar timing noise, pulse broadening owing to scattering in the ISM, tropospheric scintillation. 
Recent improvements in pulsar timing techniques \citep{lyne:2010:smr} suggest that intrinsic pulsar timing noise need not limit timing precision measurements in either present or future instrumentation. 
Unresolved pulse broadening owing to scattering in the ISM will limit the attainable $\sigma_{\tau}$ and, thus, $\sigma_L$. For nearby ($L<1$~kpc) pulsars the \emph{measured} pulse broadening at $\nu=1$~GHz is typically in the 1--10~ns range \citep{manchester:2005:atn}. Since pulse broadening is scales with observation frequency as $\nu^{-4.4}$ (though observations suggest the index may be closer to $-3.9$ \citep{bhat:2004:moo}) scattering in the ISM also need not limit our ability to measure the distance to nearby pulsars. 
Finally, recent estimates by \citet{jenet:2010:pts} indicate that, for gravitational wave frequencies $\lesssim\,\text{yr}^{-1}$, tropospheric scintillation will contribute to the noise budget at no more than the 0.03~ns rms level. 
Consequently, over the next decade $\sigma_L\lesssim1\,\text{pc}$ is thus a reasonable expectation for millisecond pulsars at distances up to a kpc. 

\subsection{Correction to plane wave approximation}
Turn now to the ratio of Fourier amplitudes
\begin{align}
\rho(f,\hat{k}_0) &= 
\frac{\widetilde{\tau}_{\text{cr}}}{\widetilde{\tau}_{\text{pw}}}\nonumber\\
&= 
\exp\left[
\pi ifL\left(2\sin^2\frac{\theta}{2}+\frac{\epsilon}{2}\sin^2\theta\right)
\right]\nonumber\\
&\qquad{}\times
\frac{
\sin\left[\frac{1}{2}\pi fL\epsilon\sin^2\theta\right]
}{
\sin\left(2\pi fL\sin^2\frac{\theta}{2}\right)
}. \label{eq:rho}
\end{align}
When $\pi fL\epsilon\sin^2\theta\geq1$ each of the $\sin$ functions whose ratio determines the $|\rho|$ has order-unity magnitude and $|\rho|$ oscillates rapidly from $0$ to $\infty$. 
For a typical PTA pulsar (distance $L\simeq$~kpc) this will be the case for source frequencies greater than or of order $\text{yr}^{-1}$ and distances $R$ less than or of order 100~Mpc (cf.\ Eqs.~\ref{eq:numerology}). In this regime the gravitational wave phasefront curvature plays an essential role in determining the response of a pulsar timing array. 

As an example consider the residuals in the pulse arrival times for PSR J$1939+2134$ owing to gravitational waves from a supermassive black hole binary system like that proposed by \citet{sudou:2003:omi} in the radio galaxy 3C~66B  at RA 2h23m11.4s, Dec $42^{\text{o}}$59'31'' and luminosity  distance $R=85.8$~Mpc \citep{ned:2010:ned}. The binary, ruled-out by the subsequent analysis of \citet{jenet:2004:cpo}, was supposed to have a total mass of $5.4\times10^{10}\mathrm{M_{\odot}}$, a mass ratio of 0.1, and a period of 1.05~yr, corresponding to a gravitational wave frequency of $1.9\,\text{yr}^{-1}$. Assuming these binary parameters to be exact and taking the distance and location of J$1939+2134$ as provided by the ATNF pulsar catalog \citep{manchester:2005:atn}  yields  $\rho=1.1\exp[-0.86\pi i]$: i.e., the magnitude of the ``correction'' $\tau_{\text{cr}}$ is greater than the ``leading-order'' plane-wave approximation contribution, the magnitude of $\tau_{\text{gw}}$ is 46\% of that predicted by the plane wave approximation, and the phase of $\tau_{\text{gw}}$ is retarded by $\pi/2$ radians relative to $\tau_{\text{pw}}$.\footnote{This comparison is meant only to illustrate the importance of $\tau_{\text{cr}}$ in estimating $\tau_{\text{gw}}$. In this case the distance to J$1939+2134$ is not known accurately enough allow a prediction for an accurate estimate of either $\tau_{\text{pw}}$ or $\tau_{\text{gw}}$.} 

\subsection{Gravitational wave source distance and location on sky}\label{sec:distances}
The magnitude and phase of the plane-wave approximation Fourier coefficient function $\widetilde{\tau}_{\text{pw}}$ depends on the distance to the associated pulsar and the pulsar-Earth-source angle. The correction depends on these and, in addition, the source distance. Measurements of $\widetilde{\tau}_{\text{gw}}$ in a collection of pulsars whose angular location and distance are known sufficiently accurately can thus be used to measure the curvature of the gravitational radiation phase-fronts and, thus, the luminosity distance and direction to the gravitational wave source. 

Consider a periodic source of gravitational waves observed in an array of pulsars whose distances and relative locations are known to high precision. The gravitational wave contribution to the pulse arrival times for array pulsar $\ell$ may be represented by an amplitude and phase, which will depend on the different $\epsilon_\ell=L_\ell/R$, $\pi fL_\ell$ and $\theta$ for each pulsar $\ell$ in the array. Requiring that these phases and amplitudes all be consistent is a powerful constraint on the source distance and location on the celestial sphere.

\begin{figure}
\includegraphics[width=3.5in]{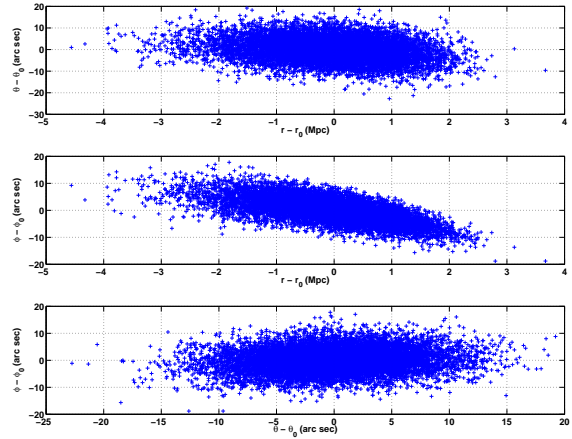}
\caption{Scatter plots of the timing parallax estimates of the distance, declination and right ascension of a simulated gravitational wave source in 3C~66B (i.e., $r=87.8$~Mpc, dec=$2^{\text{o}}59'31''$ and right ascension 2h23m11). See \S\ref{sec:distances} for details.}\label{fig:pp}
\end{figure}

As a crude but effective proof-of-principle demonstration of distance measurement by gravitational wave timing parallax consider observations made with a selection of IPTA pulsars with distances whose precision has been projected, as described above, into the SKA era. Focus attention on the example source in 3C~66B. For this source $\left(\pi f\right)^{-1} = 0.051\,\text{pc}$. Following Equation \ref{eq:sigmaL}, four pulsars currently monitored as part of the International Pulsar Timing Array  
are close enough that we may expect their distance to be measured with a one-sigma uncertainty of less than 0.025~pc.  Table \ref{tbl:ipta} lists these pulsars and their currently measured distances. For these four pulsars,
\begin{enumerate}
\item Calculate the four $\widetilde{\tau}_{\text{gw}}$ for the example source in 3C~66B described above. These complex amplitudes, corresponding to periodic timing residual amplitude and phase, are our ``observations''.  Denote the $\widetilde{\tau}_{\text{gw}}$ for pulsar $k$ by $\widetilde{\tau}_k$.\label{step:one} 
\item Adopt approximate distances to each of these four pulsars consistent with normally-distributed measurement error with standard deviation as given in Equation \ref{eq:sigmaL}. \label{step:two}
\item Using these approximate distances evaluate 
\begin{align}
\psi^2(r) &= \sum_k \left|\log\left(\frac{\widetilde{\tau}'_{k}(r)}{\widetilde{\tau}_{k}}\right)\right|^2
\end{align}
where $\widetilde{\tau}'_k(r)$ is the expected $\widetilde{\tau}_{\text{gw}}$ for pulsar $k$ \emph{assuming the source at distance $r$ and pulsar $k$ at the approximate distance found in step} \ref{step:two}.
\end{enumerate}
The quantity $\psi^2(r)$ is a measure of the misfit between the observations $\widetilde{\tau}_k$ and the prediction $\widetilde{\tau}'_k$ assuming pulsars at the approximate distances and source at distance $r$. The $r$ that minimizes $\psi^2(r)$ is thus an estimator for the distance to the source. Figure \ref{fig:pp} shows a set of three scatter plots of the estimated $r$, declination $\theta$, and right ascension $\phi$ relative to their actual values over $10^4$ realizations of pulsar distance errors, for this example. Table \ref{tbl:stats} provides quantitative descriptive statistics for the distribution of errors. Clearly, if the pulsar distances can be sufficiently accurately determined, the distance and location of a detected periodic gravitational wave source can be accurately measured. 

\begin{table*}
\caption{Four nearby millisecond pulsars monitored as part of the International Pulsar Timing Array and their current estimated distances. These four pulsars are close enough that, in the SKA era, their distances, or the distances to pulsars like them, may become known to better than 0.025~pc.}\label{tbl:ipta}
\begin{tabular}{cr}
\hline
J-Name&{kpc}\\
\hline
J$0437-4715$ & 0.16 \\
J$1741+1300$ & 0.19 \\
J$0030+0451$ & 0.30 \\
J$2124-3358$ & 0.25 \\
\hline
\end{tabular}
\end{table*}

\begin{table*}
\caption{Mean, median and standard deviation of the timing parallax estimated distance and location on the sky of a simulated gravitational wave source in 3C~66B. In this example the source was actually located at $r=87.8$~Mpc, dec=$2^{\text{o}}59'31''$ and right ascension 2h23m11s. Only the four pulsars described in Table \ref{tbl:ipta} were used to make these estimates. See \S\ref{sec:distances} for details.}\label{tbl:stats}
\begin{tabular}{lrrr}
\hline
&mean&median&std.\ dev.\\
\hline
$r$&85.72~Mpc&85.79~Mpc&0.99~Mpc\\
$\theta$&$42^{\text{o}}59'31''$&$42^{\text{o}}59'31''$&$0^{\text{o}}0'5.3''$\\
$\phi$&2h23m11.0s&2h23m11.0s&0h0m0.3s\\
\hline
\end{tabular}
\end{table*}

We emphasize that this is no more than a proof-of-principle demonstration. It is naive in its treatment of uncertainties in the $L_k$, ignores pulsar timing noise in the measured  $\widetilde{\tau}_k$, presumes a sufficiently accurate knowledge of pulsar locations $\theta_k$, and makes use of an ad hoc misfit statistic $\psi^2(r)$. Nevertheless, taken as a proof-of-principle demonstration it shows that the potential exists for pulsar timing array observations to determine the distance to periodic gravitational wave sources if the pulse arrival time dependence on the passing waves is properly modeled. 

\section{Conclusion}\label{sec:concl}

Gravitational waves crossing the pulsar-Earth line-of-sight lead to a disturbance in the pulsar pulse arrival time. All previous derivations of this disturbance have assumed planar gravitational wave phasefronts. The gravitational wave phasefronts from point gravitational wave sources --- e.g., coalescing supermassive black hole binary systems --- are curved, with curvature radius equal to the source luminosity distance. The approximation of planar wavefronts is thus valid only at distances $R\gg2\pi fL^2/c$, where $R$ is the Earth-source distance, $L$ the Earth-pulsar distance, and $f$ the gravitational radiation frequency. For typical pulsars distances (kpc) and relevant gravitational wave frequencies ($f\lesssim\,\text{yr}^{-1}$) the correction to the disturbance magnitude and phase owing to phasefront curvature is thus significant for sources within $\sim100\,\text{Mpc}$. Here we have derived the curved wavefront corrections to the pulsar timing response for such ``nearby'' sources, described when they are important, and shown that future gravitational wave observations using pulsar timing arrays may be capable of measuring luminosity distances to supermassive black hole binary systems, and other periodic gravitational wave sources, approaching or exceeding 100~Mpc. 

The gravitational wave source distance measurement described here is properly considered a parallax distance measurement. The baselines over which the parallax is measured are, in this case, the timing array pulsar-Earth baselines. Crucial to the ability to observe the effects of gravitational wave phasefront curvature is a knowledge of the pulsar-Earth distance $L$ with an uncertainty $\sigma_L\lesssim\left(\pi f\right)^{-1}\sim2\,\text{pc}\,(0.1\text{yr}^{-1}/f)$. We argue that pulsar distance measurements of this accuracy, while beyond present capabilities (except in the case of J$0437-4715$, which is exceptionally bright and close), are within the capability of SKA-era observations for pulsars at distances $L\lesssim\,\mathrm{kpc}$. 

When considered against the cosmic distance ladder, the distance measurement described here involves three rungs to reach distances greater than tens of Mpc: first, the determination of the astronomical unit; second, the distance to the array pulsars; finally, the distance to the gravitational wave source. Since the method described here also provides precise source angular position the likelihood of identifying an electromagnetic counterpart (e.g., host galaxy) to the gravitational wave source is great, raising the possibility of a precise measurement of both the redshift and luminosity distance to a single object at cosmological distances, with the obvious consequences for independent verification of the parameters describing our expanding universe. 

Even in the absence of an electromagnetic counterpart measurement of the luminosity distance and angular location of a supermassive black hole binary system --- the most likely source of periodic gravitational waves --- enables the determination of the system's so-called ''chirp mass'', or $(m_1 m_2)^{3/5}/(m_1+m_2)^{1/5}$, without the need to observe the binary's evolution. 

We have only begun to plumb the potential of gravitational wave observations as a tool of astronomical discovery. While this potential will not be realized until gravitational waves are detected and observations become, if not routine, more than occasional, explorations like these will position us to more readily exploit the opportunities that future observations present.

\section*{Acknowledgments}
We gratefully acknowledge helpful discussions with Jim Cordes, Rick Jenet, Andrea Lommen, Duncan Lorimer, David Nice, Scott Ransom, Alberto Sesana, Dan Stinebring, and Ben Stappers. This research has made use of the ATNF Pulsar Catalogue \citep{manchester:2005:atn} and the NASA/IPAC Extragalactic Database (NED). NED is operated by the Jet Propulsion Laboratory, California Institute of Technology, under contract with the National Aeronautics and Space Administration. This work was supported in part by National Science Foundation grants PHY 06-53462 (LSF) and the Penn State Physics Department (XD).

\label{lastpage}

\begin{thebibliography}{}

\bibitem[\protect\citeauthoryear{{Bhat}, {Cordes}, {Camilo}, {Nice} \&
  {Lorimer}}{{Bhat} et~al.}{2004}]{bhat:2004:moo}
{Bhat} N.~D.~R.,  {Cordes} J.~M.,  {Camilo} F.,  {Nice} D.~J.,    {Lorimer}
  D.~R.,  2004, ApJ, 605, 759

\bibitem[\protect\citeauthoryear{{Cordes}, {Arzoumanian}, {Brisken}, {Freire},
  {Kramer}, {Lai}, {Lasio}, {McLaughlin}, {Nice}, {Stairs} \&
  {Weisberg}}{{Cordes} et~al.}{2009}]{cordes:2009:tog}
{Cordes} J.,  {Arzoumanian} Z.,  {Brisken} W.,  {Freire} P.,  {Kramer} M.,
  {Lai} D.,  {Lasio} J.,  {McLaughlin} M.,  {Nice} D.,  {Stairs} I.,
  {Weisberg} J.,  2009, in astro2010: The Astronomy and Astrophysics Decadal
  Survey Vol.~2010, Tests of gravity and neutron star properties from precision
  pulsar timing and interferometry.
pp 56--+

\bibitem[\protect\citeauthoryear{Damour \& Vilenkin}{Damour \&
  Vilenkin}{2001}]{damour:2001:gwb}
Damour T.,  Vilenkin A.,  2001, Phys. Rev. D, 64, 064008

\bibitem[\protect\citeauthoryear{{Deller}}{{Deller}}{2009}]{deller:2009:pva}
{Deller} A.~T.,  2009, ArXiv e-prints, 0902.1000

\bibitem[\protect\citeauthoryear{{Deller}, {Verbiest}, {Tingay} \&
  {Bailes}}{{Deller} et~al.}{2008}]{deller:2008:ehp}
{Deller} A.~T.,  {Verbiest} J.~P.~W.,  {Tingay} S.~J.,    {Bailes} M.,  2008,
  ApJL, 685, L67

\bibitem[\protect\citeauthoryear{{Detweiler}}{{Detweiler}}{1979}]{detweiler:19%
79:ptm}
{Detweiler} S.,  1979, ApJ, 234, 1100

\bibitem[\protect\citeauthoryear{Finn}{Finn}{1985}]{finn:1985:gwf}
Finn L.~S.,  1985, Class. Quantum Grav., 2, 381

\bibitem[\protect\citeauthoryear{Finn}{Finn}{2009}]{finn:2009:roi}
Finn L.~S.,  2009, Phys. Rev. D, 79

\bibitem[\protect\citeauthoryear{{Finn} \& {Lommen}}{{Finn} \&
  {Lommen}}{2010}]{finn:2010:dla}
{Finn} L.~S.,  {Lommen} A.~N.,  2010, ApJ, 718, 1400

\bibitem[\protect\citeauthoryear{Finn \& Sutton}{Finn \&
  Sutton}{2002}]{finn:2002:bmo}
Finn L.~S.,  Sutton P.~J.,  2002, Phys. Rev. D, 65, 044022

\bibitem[\protect\citeauthoryear{Foster \& Backer}{Foster \&
  Backer}{1990}]{foster:1990:cpt}
Foster R.~S.,  Backer D.~C.,  1990, ApJ, 361, 300

\bibitem[\protect\citeauthoryear{Jenet, Armstrong \& Tinto}{Jenet
  et~al.}{2010}]{jenet:2010:pts}
Jenet F.~A.,  Armstrong J.~W.,    Tinto M.,  2010, Pulsar Timing Sensitivity to
  Very-Low-Frequency Gravitational Waves, In preparation

\bibitem[\protect\citeauthoryear{{Jenet}, {Hobbs}, {van Straten}, {Manchester},
  {Bailes}, {Verbiest}, {Edwards}, {Hotan}, {Sarkissian} \& {Ord}}{{Jenet}
  et~al.}{2006}]{jenet:2006:ubo}
{Jenet} F.~A.,  {Hobbs} G.~B.,  {van Straten} W.,  {Manchester} R.~N.,
  {Bailes} M.,  {Verbiest} J.~P.~W.,  {Edwards} R.~T.,  {Hotan} A.~W.,
  {Sarkissian} J.~M.,    {Ord} S.~M.,  2006, ApJ, 653, 1571

\bibitem[\protect\citeauthoryear{Jenet, Lommen, Larson \& Wen}{Jenet
  et~al.}{2004}]{jenet:2004:cpo}
Jenet F.~A.,  Lommen A.,  Larson S.~L.,    Wen L.,  2004, ApJ, 606, 799

\bibitem[\protect\citeauthoryear{Kaspi, Taylor \& Ryba}{Kaspi
  et~al.}{1994}]{kaspi:1994:hto}
Kaspi V.~M.,  Taylor J.~H.,    Ryba M.,  1994, ApJ, 428, 713

\bibitem[\protect\citeauthoryear{Leblond, Shlaer \& Siemens}{Leblond
  et~al.}{2009}]{leblond:2009:gwf}
Leblond L.,  Shlaer B.,    Siemens X.,  2009, Phys. Rev. D, 79, 123519

\bibitem[\protect\citeauthoryear{Lommen}{Lommen}{2001}]{lommen:2001:pmt}
Lommen A.~N.,  2001, PhD thesis, University of California, Berkeley, Berkeley,
  CA, U.S.A.

\bibitem[\protect\citeauthoryear{{Lommen}}{{Lommen}}{2002}]{lommen:2002:nlo}
{Lommen} A.~N.,  2002, in {Becker} W.,  {Lesch} H.,   {Tr{\"u}mper} J.,  eds,
  Neutron Stars, Pulsars, and Supernova Remnants New limits on gravitational
  radiation using pulsars.
pp 114--+

\bibitem[\protect\citeauthoryear{{Lommen} \& {Backer}}{{Lommen} \&
  {Backer}}{2001}]{lommen:2001:upt}
{Lommen} A.~N.,  {Backer} D.~C.,  2001, ApJ, 562, 297

\bibitem[\protect\citeauthoryear{Lyne, Hobbs, Kramer, Stairs \& Stappers}{Lyne
  et~al.}{2010}]{lyne:2010:smr}
Lyne A.,  Hobbs G.,  Kramer M.,  Stairs I.,    Stappers B.,  2010, Science,
  329, 408

\bibitem[\protect\citeauthoryear{{Manchester}, {Hobbs}, {Teoh} \&
  {Hobbs}}{{Manchester} et~al.}{2005}]{manchester:2005:atn}
{Manchester} R.~N.,  {Hobbs} G.~B.,  {Teoh} A.,    {Hobbs} M.,  2005, AJ, 129,
  1993

\bibitem[\protect\citeauthoryear{NASA/JPL}{NASA/JPL}{2010}]{ned:2010:ned}
NASA/JPL, 2010, NASA/IPAC Extragalactic Database

\bibitem[\protect\citeauthoryear{{Pshirkov}, {Baskaran} \&
  {Postnov}}{{Pshirkov} et~al.}{2010}]{pshirkov:2010:ogw}
{Pshirkov} M.~S.,  {Baskaran} D.,    {Postnov} K.~A.,  2010, {MNRAS}, 402, 417

\bibitem[\protect\citeauthoryear{Romani \& Taylor}{Romani \&
  Taylor}{1983}]{romani:1983:ulo}
Romani R.~W.,  Taylor J.~H.,  1983, ApJL, 265, L35

\bibitem[\protect\citeauthoryear{{Sazhin}}{{Sazhin}}{1978}]{sazhin:1978:ofd}
{Sazhin} M.~V.,  1978, Soviet Astronomy, 22, 36

\bibitem[\protect\citeauthoryear{Sesana}{Sesana}{2010}]{sesana:2010:scm}
Sesana A.,  2010, arXiv, 1006.0730v1

\bibitem[\protect\citeauthoryear{{Sesana} \& {Vecchio}}{{Sesana} \&
  {Vecchio}}{2010}]{sesana:2010:gwa}
{Sesana} A.,  {Vecchio} A.,  2010, Class. Quantum Grav., 27, 084016

\bibitem[\protect\citeauthoryear{{Seto}}{{Seto}}{2009}]{seto:2009:sfm}
{Seto} N.,  2009, {MNRAS}, 400, L38

\bibitem[\protect\citeauthoryear{{Siemens}, {Mandic} \& {Creighton}}{{Siemens}
  et~al.}{2007}]{siemens:2007:gsb}
{Siemens} X.,  {Mandic} V.,    {Creighton} J.,  2007, Phys. Rev. Lett., 98,
  111101

\bibitem[\protect\citeauthoryear{{Smits}, {Kramer}, {Stappers}, {Lorimer},
  {Cordes} \& {Faulkner}}{{Smits} et~al.}{2009}]{smits:2009:psa}
{Smits} R.,  {Kramer} M.,  {Stappers} B.,  {Lorimer} D.~R.,  {Cordes} J.,
  {Faulkner} A.,  2009, A\&A, 493, 1161

\bibitem[\protect\citeauthoryear{{Sudou}, {Iguchi}, {Murata} \&
  {Taniguchi}}{{Sudou} et~al.}{2003}]{sudou:2003:omi}
{Sudou} H.,  {Iguchi} S.,  {Murata} Y.,    {Taniguchi} Y.,  2003, Science, 300,
  1263

\bibitem[\protect\citeauthoryear{Sutton \& Finn}{Sutton \&
  Finn}{2002}]{sutton:2002:bgm}
Sutton P.~J.,  Finn L.~S.,  2002, Class. Quantum Grav., 19, 1355

\bibitem[\protect\citeauthoryear{Taylor, Fowler \& McCulloch}{Taylor
  et~al.}{1979}]{taylor:1979:mog}
Taylor J.~H.,  Fowler L.~A.,    McCulloch P.~M.,  1979, Nature, 277, 437

\bibitem[\protect\citeauthoryear{Tingray, Bignall, Colegate, Aben, Nicolls \&
  Weston}{Tingray et~al.}{2010}]{tingray:2010:har}
Tingray S.,  Bignall H.,  Colegate T.,  Aben G.,  Nicolls J.,    Weston S.,
  2010, in Proceedings of SKA2010 - International SKA Science and Engineering
  Meeting The high angular resolution component of the {SKA}.
University of Manchester, UK

\bibitem[\protect\citeauthoryear{{van Haasteren} \& {Levin}}{{van Haasteren} \&
  {Levin}}{2010}]{van-haasteren:2010:gma}
{van Haasteren} R.,  {Levin} Y.,  2010, {MNRAS}, 401, 2372

\bibitem[\protect\citeauthoryear{{Verbiest}, {Bailes}, {van Straten}, {Hobbs},
  {Edwards}, {Manchester}, {Bhat}, {Sarkissian}, {Jacoby} \&
  {Kulkarni}}{{Verbiest} et~al.}{2008}]{verbiest:2008:pto}
{Verbiest} J.~P.~W.,  {Bailes} M.,  {van Straten} W.,  {Hobbs} G.~B.,
  {Edwards} R.~T.,  {Manchester} R.~N.,  {Bhat} N.~D.~R.,  {Sarkissian} J.~M.,
  {Jacoby} B.~A.,    {Kulkarni} S.~R.,  2008, ApJ, 679, 675

\bibitem[\protect\citeauthoryear{{Weisberg} \& {Taylor}}{{Weisberg} \&
  {Taylor}}{2005}]{weisberg:2005:rbp}
{Weisberg} J.~M.,  {Taylor} J.~H.,  2005, in Rasio F.,  Stairs I.~H.,  eds,
  Binary Radio Pulsars The relativistic binary pulsar b1913+16.
Astronomical Society of the Pacific, San Francisco, pp 25--31

\bibitem[\protect\citeauthoryear{{Yardley}, {Hobbs}, {Jenet}, {Verbiest},
  {Wen}, {Manchester}, {Coles}, {van Straten}, {Bailes}, {Bhat},
  {Burke-Spolaor}, {Champion}, {Hotan} \& {Sarkissian}}{{Yardley}
  et~al.}{2010}]{yardley:2010:sop}
{Yardley} D.~R.~B.,  {Hobbs} G.~B.,  {Jenet} F.~A.,  {Verbiest} J.~P.~W.,
  {Wen} Z.~L.,  {Manchester} R.~N.,  {Coles} W.~A.,  {van Straten} W.,
  {Bailes} M.,  {Bhat} N.~D.~R.,  {Burke-Spolaor} S.,  {Champion} D.~J.,
  {Hotan} A.~W.,    {Sarkissian} J.~M.,  2010, ArXiv e-prints, 1005.1667

\end{thebibliography}
\end{document}